# Over seven decades of solar microwave data obtained with Toyokawa and Nobeyama Radio Polarimeters

Masumi Shimojo
National Astronomical Observatory of Japan, National Institutes of Natural Sciences,
Mitaka, Tokyo 181-8588, Japan
Department of Astronomical Science, The Graduate University of Advanced Studies, SOKENDAI,
Mitaka, Tokyo 181-8588, Japan
https://orcid.org/0000-0002-2350-3749
E-mail: masumi.shimojo@nao.ac.jp

Kazumasa Iwai
Institute for Space-Earth Environmental Research, Nagoya University,
Furo-cho, Chikusa-ku, Nagoya, 464-8601, Japan
https://orcid.org/0000-0002-2464-5212
E-mail: k.iwai@isee.nagoya-u.ac.jp

## Abstract

Monitoring observations of solar microwave fluxes and their polarization began in Japan during the 1950s at Toyokawa and Mitaka. At present (April 2022), monitoring observations continue with the Nobeyama Radio Polarimeters (NoRP) at the Nobeyama campus of the National Astronomical Observatory of Japan (NAOJ). In this paper, we present a brief history of the solar microwave monitoring observations preceding those now carried out by NoRP. We then review the solar microwave obtained at Toyokawa and Nobeyama and their metadata. The datasets are publicly provided by the Solar Data Archive System (SDAS) operated by the Astronomy Data Center of the NAOJ, via http (https://solar.nro.nao.ac.jp/norp/) and FTP (ftp://solar-pub.nao.ac.jp/pub/nsro/norp/) protocols.

## 1. Introduction

The Sun is a critical element of the terrestrial environment. As the range of human activities has recently expanded to include space (with, for example, commercial space travel and satellite utilization for ordinary people), the importance of predicting the near-Earth space conditions has rapidly increased. We call such space conditions "space weather." Solar activities, especially solar flares, and coronal mass ejections (CMEs), are the highest-priority phenomena for predicting space weather. Research in space weather is crucial not only for research on our environment but also for exoplanet studies. We have already found over 4800 candidates for exoplanets[1], mainly from data obtained with the Kepler satellite and the Transiting Exoplanet Survey Satellite (TESS). Exoplanet studies have progressed just detecting exoplanets to considering their habitability. The influence of the central star is essential for evaluating habitability. For example, the radiation from the central star, especially ultraviolet radiation, strongly affects the chemical evolution in the exoplanet's atmosphere (Airapetian et al., 2016). Therefore, in such studies, the Sun is a template (Shimojo 2021). As described, observations of the Sun are helpful in many fields.

Solar measurements form one of the oldest astronomical datasets. The most famous is the sunspot number, which has been measured since the observation by Thomas Harriot in the early 17th century until now (Clette et al., 2014, Muñoz-Jaramillo and Vaquero, 2019, Vokhmyanin, Arlt and Zolotova, 2020). During World War II, modern astronomical observations began to sample non-visible radiations. Radio astronomy is one of such modern astronomical observations, and bursty radio emission (radio bursts) from the Sun were detected during the development of radar in the UK (Hey 1946). From the variation in solar radio fluxes, we can easily identify transient solar activities, such as solar flares and CMEs, which are the primary sources of disturbances in near-Earth space. Solar monitoring observations with radio waves began at a few observatories in the late 1940s (Tanaka et al., 1973). The most famous and longest solar radio observation has been conducted at 2.8 GHz in Ottawa, Canada since 1947 (Covington 1948, Tapping 2013). Presently the 2.8 GHz observation contunes at the Dominion Radio Astrophysical Observatory (DRAO), Penticton BC, Canada. The radio flux measured at local noon at the DRAO is called the “F10.7 index” and is widely used as an indicator of daily solar activities not only for heliophysics studies but also for studies of the Earth's upper atmosphere: the solar radio flux is frequently used as a proxy for solar ultraviolet irradiance, which has a significant influence on the Earth's upper atmosphere (Richards et al., 1994).

In Japan, solar microwave monitoring observations began in the early 1950s. At present, the monitoring observation is carried out with the Nobeyama Radio Polarimeters (NoRP) operated by the National Astronomical Observatory of Japan (NAOJ). The NoRP is constructed from 6 equatorial-mount antennas with 8 parabolic dishes and measures the total solar flux and its circular polarization at 1, 2, 3.7, 9.4, 17, 35, and 80 GHz from dawn to dusk (Figure 1). NoRP observations with this wide frequency coverage have been extensively used to measure gyrosynchrotron emissions from solar flares (e.g., Fleishman, Bastian and Gary 2008). The observing history for 3.75 GHz extends over 70 years, and the observing history at several other frequencies (1, 2, 3.75, and 9.4 GHz) also exceeds 60 years at present. Such a long-term observations with multiple frequencies in the microwave range are unique worldwide and provide crucial data for solar cycle studies (e.g., Shimojo et al., 2017, Tapping and Morgan 2017).

[1] NASA Exoplanet Archive https://exoplanetarchive.ipac.caltech.edu (accessed on December 20, 2021)

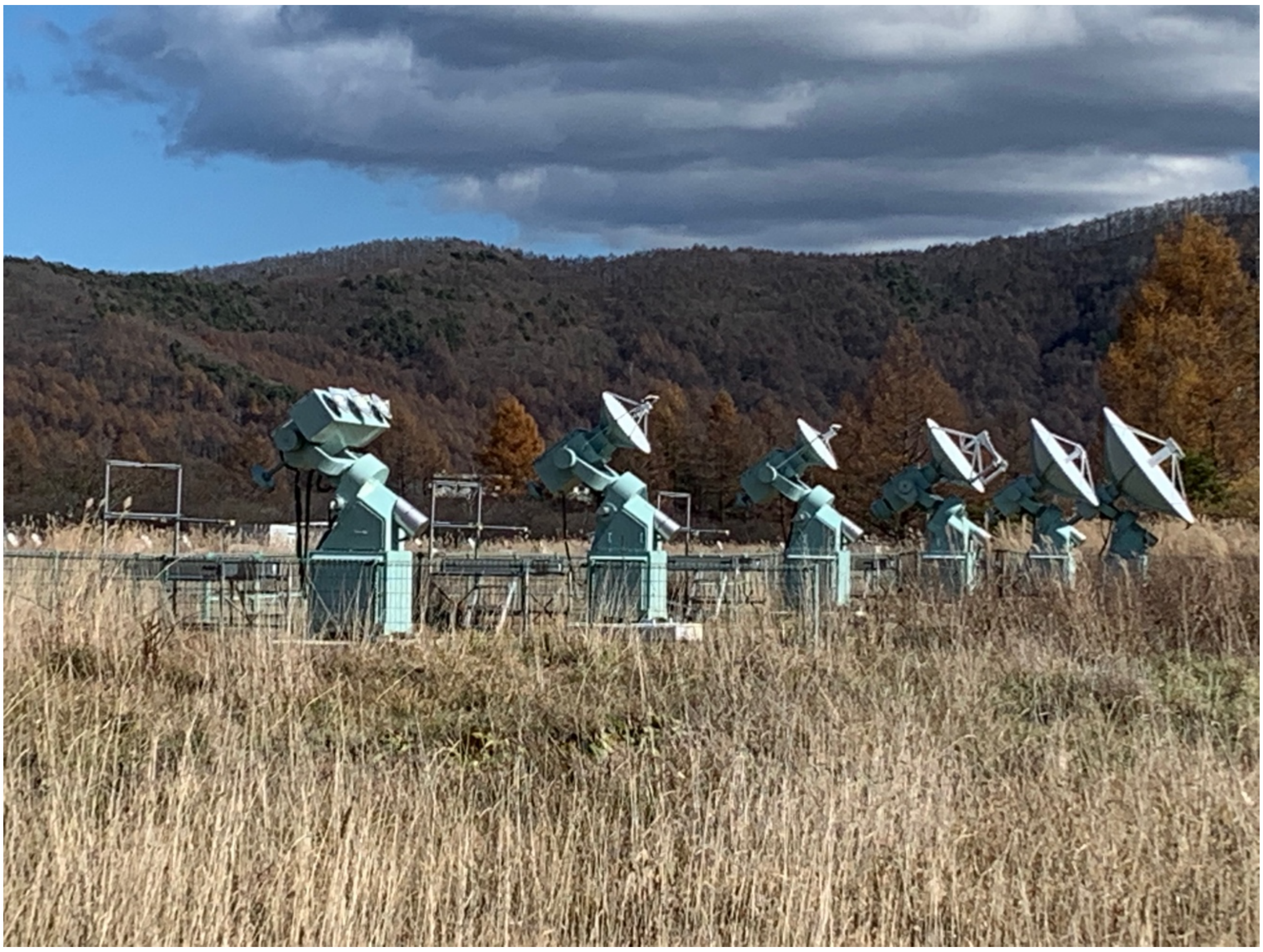

*Figure 1: A photograph of the Nobeyama Radio Polarimeters (NoRP) taken on November 10, 2021. From left, the 35/80 GHz, 17 GHz, 9.4 GHz, 3.75 GHz, 2 GHz, and 1 GHz antennas.*

The observations at all observing frequencies of the NoRP did not start at Nobeyama, and each antenna has a history. In this paper, we briefly review the history of solar microwave monitoring observations in Japan, the transitions of the NoRP instruments, and the datasets of over seven decades archived and provided by the Solar Data Archive System (SDAS) of the NAOJ. Finally, we discuss the prospects for the future of NoRP.

## 2. History of monitoring observations of solar microwave flux and polarization at Toyokawa, Mitaka, and Nobeyama

The early history of radio astronomy in Japan has been summarized in Tanaka (1984) and the series of papers "Highlighting the History of Japanese Radio Astronomy" published in the Journal of Astronomical History and Heritage (Ishiguro et al. 2012, Shimoda et al. 2013, Nakajima et al. 2014, Orchiston et al. 2016). Hence, in this paper, we briefly describe the history of monitoring observations of solar microwave fluxes and their polarization related to NoRP. Observations linked to the NoRP have been conducted at three sites (Figure 2): the Toyokawa campus (34° 50′ N, 137° 22′ E) of the Research Institute of Atmospherics, Nagoya University; the Mitaka campus (35° 40′ N, 139° 32′ E) of the Tokyo Astronomical Observatory (TAO), University of Tokyo; and the Nobeyama Solar Radio Observatory (35° 56′ N, 138° 28′ E), Tokyo Astronomical Observatory , University of Tokyo. Next, we briefly describe the history of the solar microwave monitoring observations at each site.

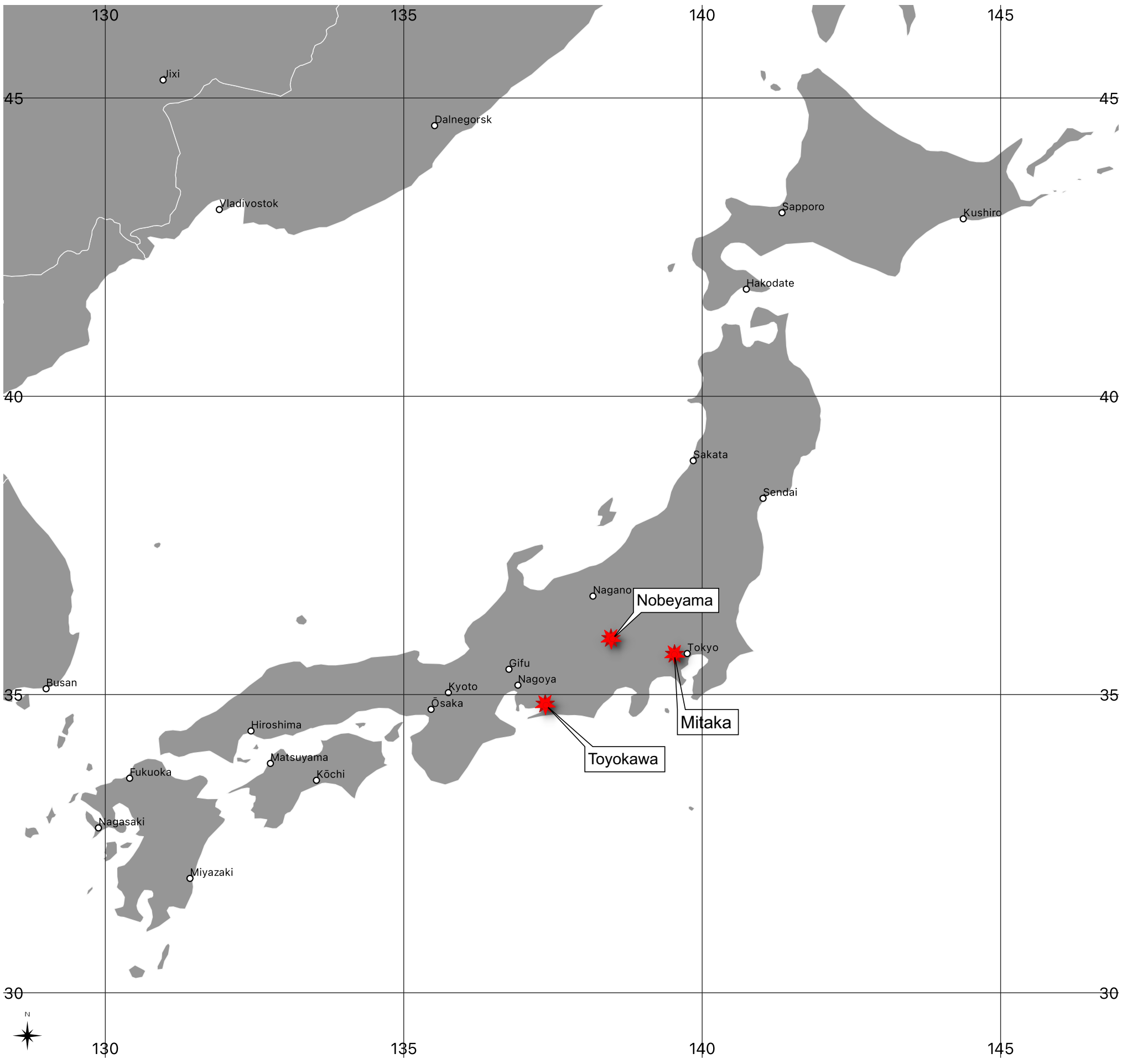

*Figure 2: Location of the sites of solar microwave monitoring in Japan.*

*The red stars indicate the Toyokawa, Mitaka, and Nobeyama sites.*

### 2.1 Toyokawa

The Research Institute of Atmospherics, Nagoya University, was established at Toyokawa in May 1949 for studying radio noise in the terrestrial atmosphere. The first relevant solar radio measurement was performed at the institute by Prof. Haruo Tanaka's team in April 1951 (Tanaka et al. 1953). They observed the total flux of the Sun at 3.75 GHz and started the monitoring observations in November 1951. Observations at 3.75 GHz have continued the present, and thus observation period at this frequency extends over 70 years.

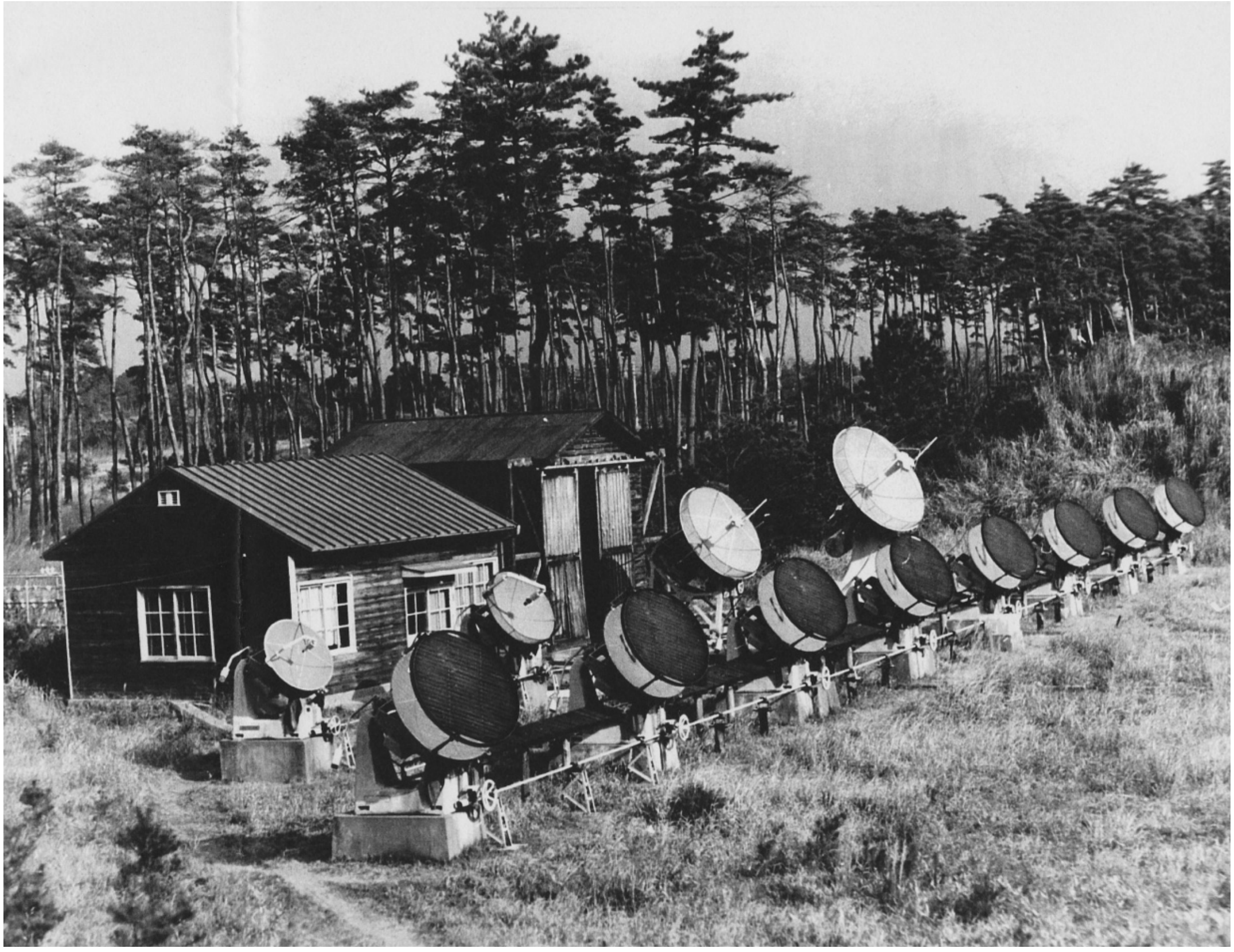

*Figure 3: Behind: Toyokawa Radio Polarimeters (ToRP) at 1 GHz, 2 GHz, 3.75 GHz and 9.4 GHz (from right) Front: the expanded 8-element solar grating array, complete with polarization screens. (Courtesy: the family of H. Tanaka)*

Prof. Tanaka and his team developed and constructed a monitoring system for solar microwave flux and polarization with multiple bands for joint observations during the International Geophysical Year (IGY: July 1957 -- December 1958). Actually, they constructed the antennas for observation at 1, 2, 9.4 GHz and started the monitoring observations in May 1956 for 9.4 GHz, May 1957 for 2 GHz, and June 1957 for 1 GHz (Tanaka et al. 1957). In this paper, we refer to the solar microwave monitoring system at 1, 2, 3.75, and 9.4 GHz as the "Toyokawa Radio Polarimeters" (ToRP: Figure 3). The first eleven years (one solar cycle) of data since the starting of the 3.75 GHz monitoring observation were summarized by Tanaka (1964).

All of the antennas constructed at Toyokawa in the 1950s were replaced in the 1970s. Based on the nameplate of the 9.4 GHz antenna, we know that the replacement year of the 9.4 GHz antenna was 1973. Unfortunately, we could not verify the replacement year of the 1 and 2 GHz antennas because their nameplates have been lost.

The high precision of the solar microwave fluxes obtained with the ToRP was recognized by the working group organized under the Commission V, International Union of Radio Science (Tanaka et al. 1973). Hence, the data have been used as the reference data for calibrating solar microwave fluxes obtained with other telescopes. This trust continues until the present, and NoRP data are also used as the reference data for new telescopes (e.g., Hwangbo et al., 2015).

The receiver systems developed in the 1950s were constructed using vacuum tubes. In 1979, the receiver systems for all observing frequencies were replaced with new systems constructed from solid-state devices (Torii et al, 1979). Due to the replacement, the stability of the receivers was improved significantly, and the operation of the telescopes was able to be fully automated. At the time, the noise generators for flux calibration were also changed from resistances to argon tubes. However, the stability of the new noise sources proved to be inadequate, and the noise generators were returned to the resistances soon (personal communication from Prof. K. Shibasaki).

The NAOJ was established in April 1988. At that time, to realize the Nobeyama Radioheliograph (NoRH), a radio interferometer dedicated to synthesizing solar images at 17 and 34 GHz (Nakajima et al. 1994), the solar radio group of the Research Institute of Atmospherics left Nagoya University, and the Nobeyama Solar Radio Observatory (NSRO) of the NAOJ was established by combining them with the Nobeyama Solar Radio Observatory of the TAO. At the beginning of the NSRO, they concentrated on the development and operation of the NoRH, while Toyokawa's staff operated the ToRP. In May 1994, because the operation of the NoRH became stable, the ToRP was moved to Nobeyama. To avoid suspending the monitoring observation at 3.75 GHz during the moving process, a new antenna for 3.75 GHz observations was constructed at Nobeyama in 1993. After completing the solar observations at Toyokawa, the old 3.75 GHz antenna was moved to the science museum "Discovery Park Yaizu" at Yaizu, Shizuoka, Japan, and it may still be seen there.

### 2.2 Mitaka

Nakajima et al. (2014) summarized the solar radio observations on the Mitaka campus of the TAO. Hence we rely here upon the history of solar monitoring observations in the microwave range from their paper.

In 1948, the radio astronomy group led by Prof. Takeo Hatanaka was established at the TAO. They constructed the first solar radio telescope at the Mitaka campus with the help of the Radio Research Laboratories of the Ministry of Posts and Telecommunications. The antenna is the origin of celestial radio observations in the TAO. A replica of the antenna was exhibited at Nobeyama.

Solar microwave observations on the Mitaka campus commenced in August 1952 with the 3 GHz observation using a 2m-dish antenna to respond to the requirements of the URSI General Assembly held in 1950. In May 1957, a 9.5 GHz polarimeter was developed, and the monitoring observations started (Akabane 1958a, 1958b). Moreover, 17 GHz monitoring observation began in May 1964.

While the monitoring observations at the Mitaka campus continued until the late 1960s, most of the data disappeared, except for the daily total solar fluxes published in the "Quarterly Bulletin on Solar Activity (QBSA)"[2] and some event data published in scientific journals (e.g., Takakura 1960, Nagasawa et al. 1961), which have been actively subjected to analyses in the latest astrophysical discussions (Cliver et al., 2020). The knowledge obtained from the observations at Mitaka was utilized for microwave observations at Nobeyama.

### 2.3 Nobeyama

Because radio frequency interference (RFI) at the Mitaka campus worsened in the 1960s, the TAO solar radio group sought a new site for solar radio observations. Given that the Nobeyama highland, Minamimaki, Nagano, Japan, is surrounded by mountains that can block radio from cities and the transportation to Nobeyama is relatively convenient[3], it was selected as the site for establishing a new solar radio observatory. In 1968, the Nobeyama Solar Radio Observatory of the TAO was established, and the 160 MHz and 17 GHz interferometers were constructed.

---

[2] https://solarwww.mtk.nao.ac.jp/en/wdc/qbsa.html

[3] Nobeyama has the train station at the highest altitude in Japan.

The altitude of the Nobeyama highland reduces the impact of the atmosphere and makes it an attractive site for higher-frequency observations. Eventually, 17, 35, and 80 GHz monitoring observations started in January 1978, May 1983, and February 1984, respectively (Nakajima et al. 1985).

In 1988, there was a major reorganization of the astronomical institutes in Japan. To construct and operate world-class telescopes, the TAO and some astronomical institutes were reorganized in the NAOJ as a national institute. At that time, Nobeyama Solar Radio Observatory (TAO) gathered with the solar radio group of the Research Institute of Atmospherics (Toyokawa), and together they established the Nobeyama Solar Radio Observatory (NSRO) of the NAOJ in order to realize the NoRH. Subsequently, a new 3.75 GHz polarimeter was constructed in Nobeyama in 1993, and the 1, 2, and 9.4 GHz polarimeters were moved from Toyokawa to Nobeyama in 1994. Thus, the NSRO of the NAOJ constructed the current NoRP comprising polarimeters at 1, 2, 3.75, 9.4, 17, 35, and 80 GHz (Figure 4).

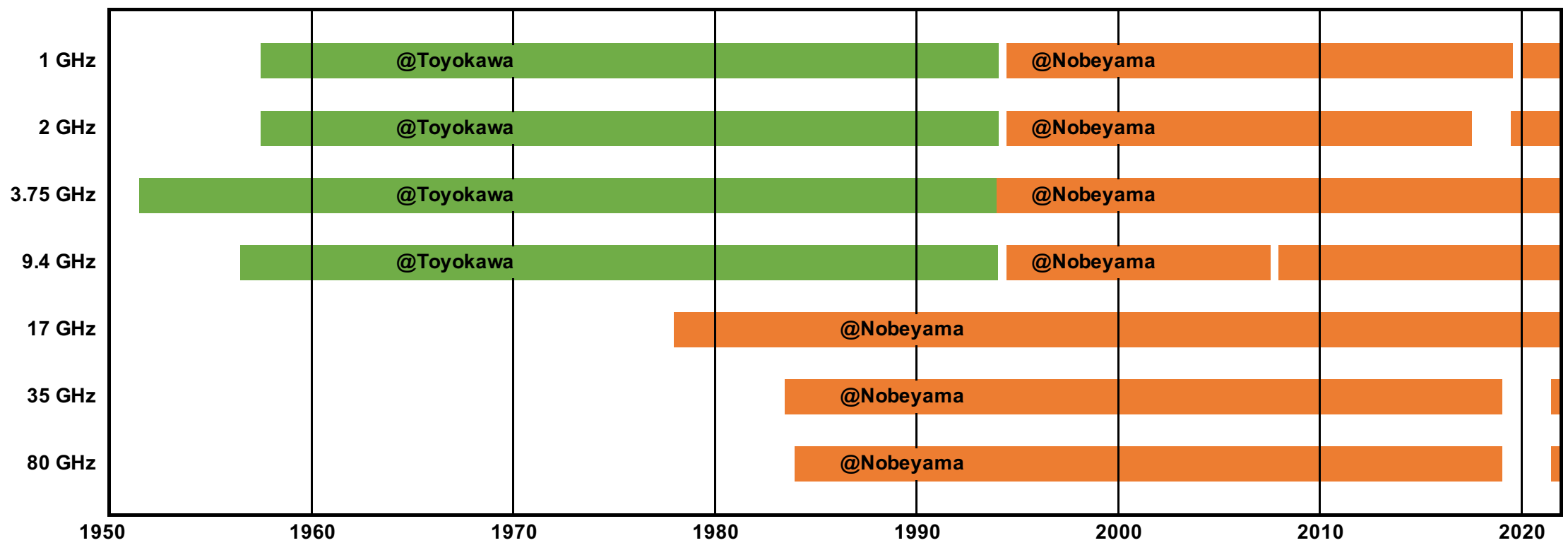

*Figure 4: The observing period of each observing frequency of the ToRP and NoRP*
*Green and orange indicate observations at Toyokawa and Nobeyama, respectively. The gaps of 1, 2, and 9.4 GHz in 1994 indicate suspensions due to the moving to Nobeyama. The other gaps show long suspensions due to trouble with the instruments and RFI.*

While the NSRO was closed at the end of March 2015, the operation of the NoRP was continued by the Nobeyama Radio Observatory, which is the division of cosmic radio astronomy at the NAOJ, until March 2019. Since April 2019, the Solar Science Observatory, the division of solar and space plasma physics at the NAOJ, has taken responsibility for the NoRP operation. As a result of the change in the division that handles the NoRP operation, it is difficult to verify the data quality from a scientific perspective because there is not always a solar radio physicist in the division. Therefore, solar radio physicists in Japan created a consortium for the NoRP scientific operation to solve this problem. The consortium holds a monthly meeting and scientifically verifies the NoRP data. In April 2022, the researchers of the following institutes have joined the consortium; Kyoto University, Ibaraki University, Nagoya University, National Institute of Information and Communications Technology, National Defense Academy and NAOJ (e.g., Shimojo et al. 2017). In April 2019, the NoRP operation was extended for a further five years with the support of the Japanese solar physics community and foreign colleagues.

## 3. Instruments of NoRP and their transitions

While the receiver systems of the NoRP have been replaced, updated, and digitized using modern instruments, the signal processing in the receivers basically follows the original designs. According to various papers on the instruments (Tanaka et al. 1953, Tanaka & Kakinuma 1957, Nakajima et al. 1985), the original receiver systems are still valid for understanding the signal processing of the NoRP. Furthermore, a calibration method for the NoRP data at 1, 2, 3.75, and 9.4 GHz was established by Prof. Tanaka and his colleagues (Tanaka et al. 1973), and we are still using their methods. For the 17, 35, and 80 GHz data, Nakajima et al. (1985) described the calibration methods for each frequency. However, RFI has been steadily increasing in the Nobeyama area, particularly since the 2000s, and this has affected the use of the original monitoring frequencies. In this section, we present the changes in the RFI in the Nobeyama area and then describe how we have dealt with it. These changes in the the NoRP instruments have not been previously reported. We acquired such information from the observing logbooks written in Japanese.

Since most internal documents of the ToRP were lost, we cannot follow the transitions of the ToRP system at Toyokawa except as described in section 2.1. Hence, this paper describes the changes that occurred after moving the ToRP to Nobeyama in May 1994.

### 3.1 Radio Frequency Interference (RFI) at the NoRP observing frequencies

Because the observing frequencies of the NoRP are not located in the protected bands for radio astronomy, we cannot avoid RFI in principle. The origins of the RFI can be roughly identified from the table of the "Frequency Assignment Plan,"[4] which is administered by the Japanese government. Houjou et al. (2010) surveyed the radio environment at Nobeyama in the 2000s and summarized the RFI and the corresponding treatments. In this section, we describe the RFI that impacts the NoRP observations based on that study and recent observation data.

### 1 GHz

The 960 -- 1164 MHz frequency range, which includes the observing frequency of the 1 GHz receiver system, is assigned to the Distance Measuring Equipment (DME), TACtical Air Navigation system for aircraft (TACAN), Air Traffic Control Radar Beacon System (ATCRBS), and Airborne Collision Avoidance System (ACAS). Hence, RFI sometimes appears NoRP data when an aircraft passes above Nobeyama, as shown in Figure 5. The radio environment survey indicated that most of the aircraft flying over Nobeyama use frequencies between 1025 and 1150 MHz.

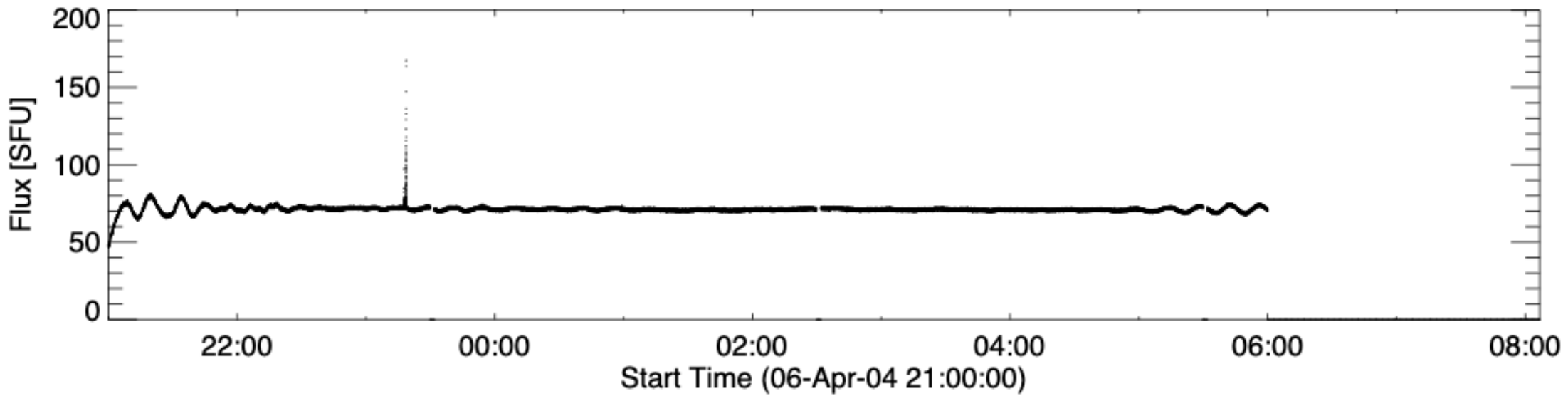

*Figure 5: The 1 GHz time profile on April 7, 2004.*
*The spike around 23:20UT indicates the interference caused by an aircraft.*

### 2 GHz

The 1980 -- 2170 MHz frequency range is assigned for communication with satellites and between cellular phones and base stations. As a result of a survey in the 2000s, Houjou et al. (2010) found that the base station for cellular phones located near the observatory uses a frequency of 2137 -- 2147 MHz. Although the frequency range is outside the nominal frequency range of the 2 GHz receiver system, it might affect the observations.

At 01:59UT on April 12, 2017, an extreme interference started as shown in Figure 6, and solar observation at 2 GHz was impossible. We measured the radio environment and found a strong signal in the range of 2110 -- 2170 MHz. We did not identify the origin of the signal, but it may have been caused by a new base station for cellular phones.

[4] https://www.tele.soumu.go.jp/e/index.htm

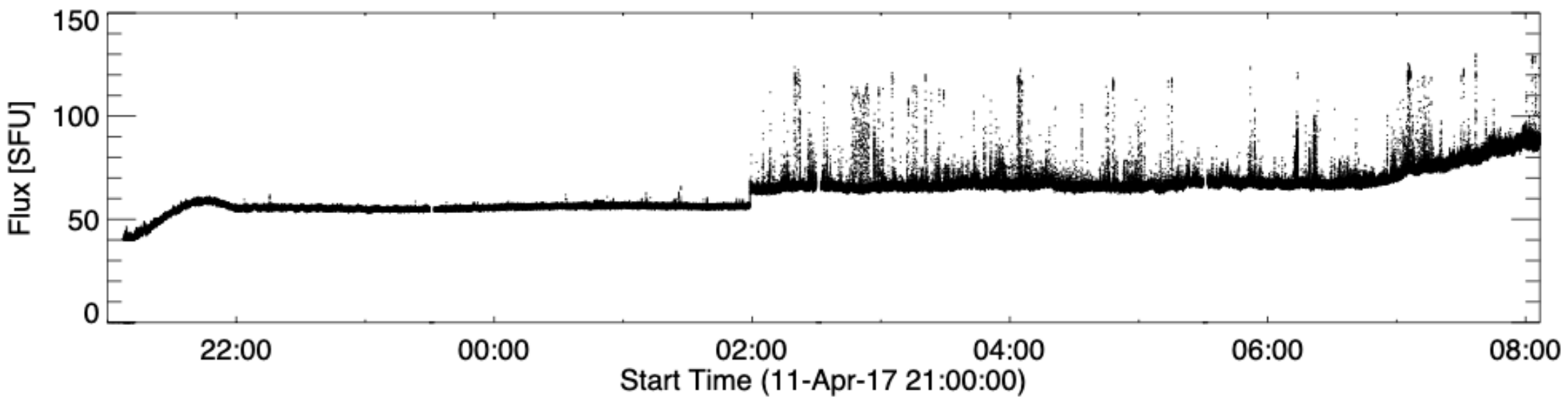

*Figure 6: The 2 GHz time profile on April 12, 2017.*

3.75 GHz

A frequency range of 3400 -- 4200 MHz is assigned for the communication with satellites. The interference by satellite communications strongly influences the 3.75 GHz observations around the spring and autumn equinoxes because the line-of-sight to the Sun passes through the geosynchronous satellite belt in this period, and the primary beam of the 3.75 GHz antenna is large enough to include a number of satellites. Figure 7 shows the 3.75 GHz data on a day near the autumn equinox, and each peak in the plot show a different geosynchronous satellite.

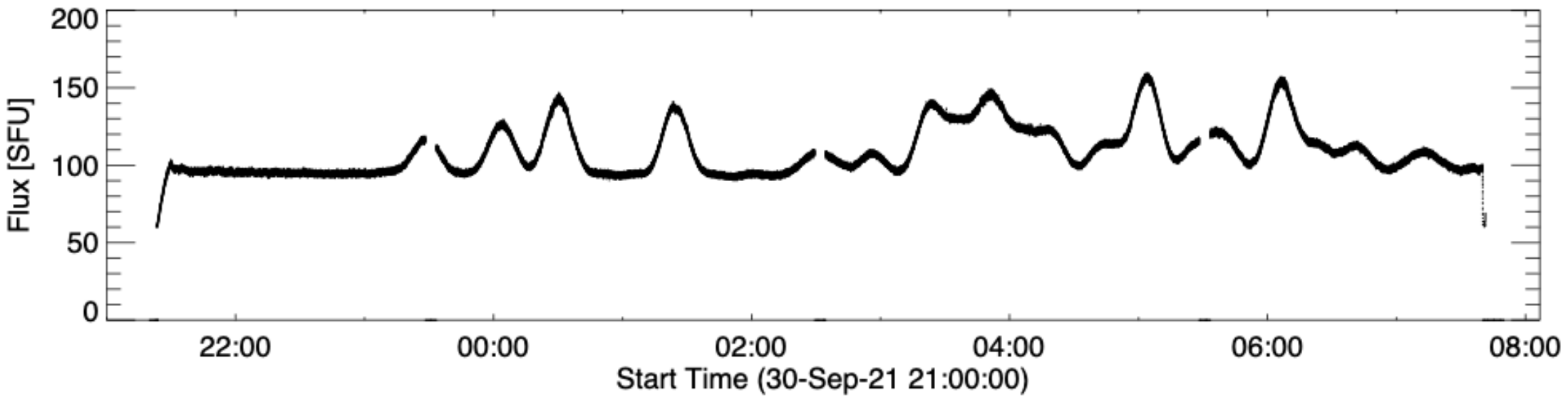

*Figure 7: The 3.75 GHz time profile on October 1, 2021.*

9.4 GHz

The 9200 -- 9800 MHz frequency range is mainly assigned to aeronautical, weather, and marine radars. Houjou et al. (2010) detected RFI only when observing the southern sky. The Yaizu fishing port, one of the largest home ports in Japan for pelagic fishing boats, is located 120 km south of Nobeyama. Hence, the interference would be caused by the powerful marine radars of pelagic fishing ships entering the Yaizu port. A spectrum survey found the interference concentrates in 9345 -- 9400 MHz and 9460 -- 9600 MHz.

Since the late 2010s, we have frequently detected RFI in the 9.4 GHz data. Interference appears once at approcimately 00UT and around 04UT on clear days (upper panel, Figure 8). Moreover, the interference has been strongly correlated with bad weather. For example, the lower panel of Figure 8 shows the 9.4 GHz time profile when a typhoon passed near Nobeyama. Hence, the interference is caused by the weather radars. Recently, local governments in Japan used weather radars for disaster prevention. We confirmed that there is no weather radar in Minamimaki village, but there are weather radars around Nobeyama.

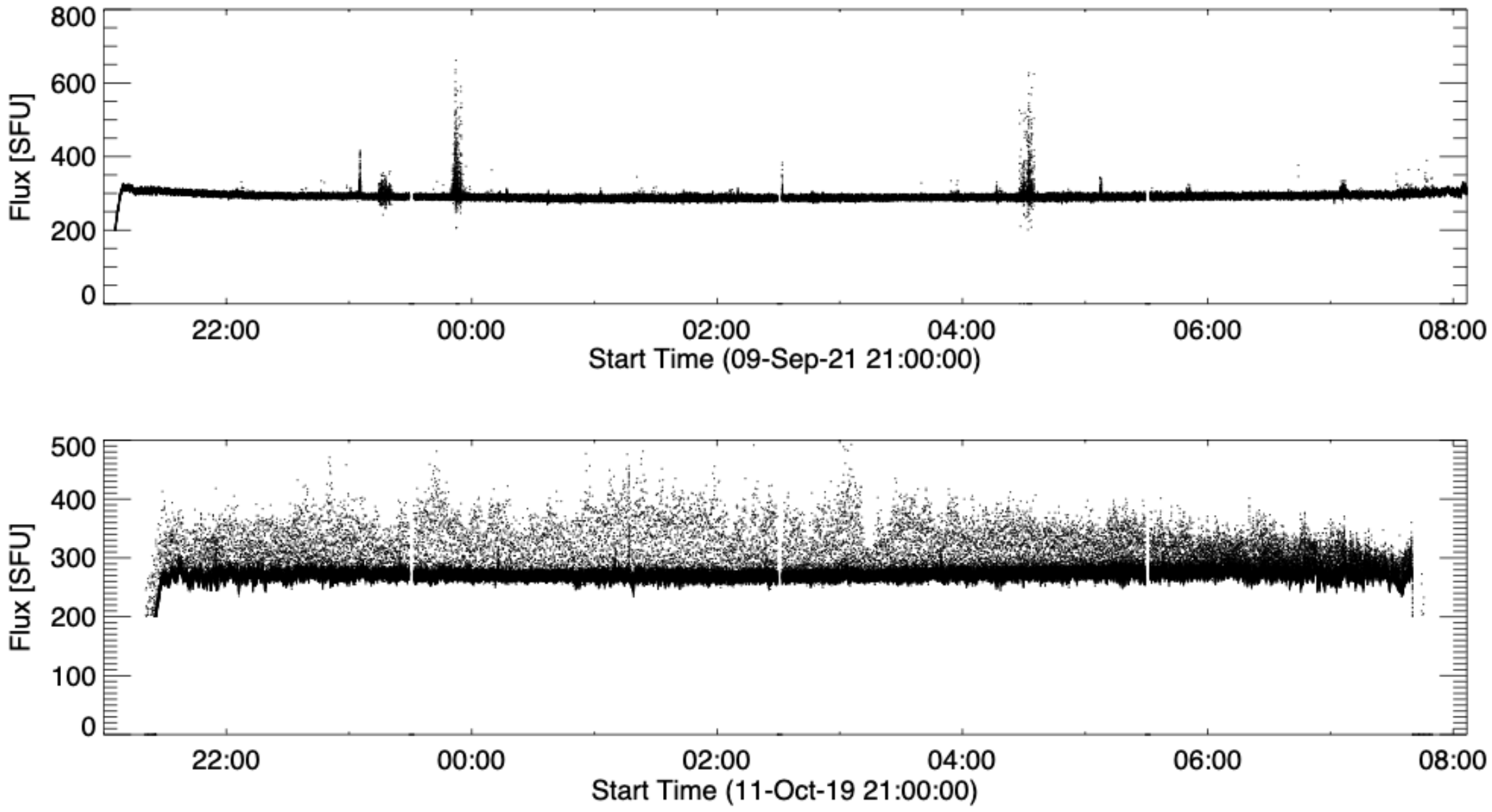

*Figure 8: The 9.4GHz time profiles*

*Upper panel: September 10, 2021 (clear day), Lower panel: October 12, 2019 (typhoon day)*

At frequencies higher than 9.4 GHz, we have yet to detect serious RFI impacts on the NoRP observations. Hence, we do not consider the RFI for the 17, 34, and 80 GHz observations. Nevertheless, the frequency range used for human activities is becoming wider. For example, millimeter waves (>30GHz) have recently been used for in-vehicle radars. We will need to consider RFI at such higher frequencies in the near future.

### 3.2 Receiver system transitions of NoRP and avoidance of RFI

As mentioned in the previous section, the observing frequencies of the NoRP are not located in the protected bands for radio astronomy. Hence, based on the radio environment survey, changing the observing frequency is the only way to avoid RFI. Before describing the details of our treatments, we briefly describe how to the observing frequencies of the NoRP receivers are set. This is usually well described and in detail in radio astronomy textbooks.

The receiver system of the NoRP is a simple heterodyne receiver with a Dicke switching function, except for the 80 GHz system. In the NoRP receiver system, the signal from the Sun collected with the parabolic antenna is mixed with the signal from the local oscillator, and then we obtain only the down-converted signal (intermediate Frequency: IF) using the frequency filter. The process is carried out in the front-end system immedeiately behind the parabolic dish.

The IF signal is sent to the back-end system located in the building. In the back-end system, a reference signal generated in the front-end system is subtracted from the solar signal (Dicke switching), and then the power of the subtracted signal is measured by a square-law detector.

The actual center frequency of the observing frequency range "$f_{RF}$" is determined from the frequency of the local oscillator "$f_{Local}$", and the IF frequency "$f_{IF}$". The two sidebands resulting from the mixing process have central frequencies $f_{RF} = f_{Local} \pm f_{IF}$. The frequency band around "$f_{Local}$-$f_{IF}$" is called the "lower side band," or "LSB". On the other hand, the frequency band around "$f_{Local}$+$f_{IF}$" is called the "upper side band," or "USB." Because the IF signal includes both LSB and USB signals, the square-law detector measures the power of the LSB and USB combined. We call such a receiver system the "double side band" or "DSB" receiver. With a frequency filter inserted before mixing the signal from the local oscillator, we can select either the LSB or USB included in the IF signal. In this case, we call this a "single side band" or "SSB" receiver. The $f_{IF}$ of the NoRP is fixed to 60 MHz for the 1, 2, 3.75, and 9.4 GHz. Hence, we can change the observing frequency by changing $f_{Local}$ and/or putting a frequency filter before the mixer.

The observing frequencies of the ToRP as described by Torii et al. (1979), the newest paper describing ToRP, and those when the ToRP arrived at Nobeyama in 1994 are presented in Table 1. All the ToRP receiver systems were originally the DSB receivers. The $f_{Local}$ of 3.75 and 9.4 GHz at Nobeyama were already different from those described by Torii et al. (1979), and the 3.75 GHz receiver system was changed from a DSB receiver to an SSB receiver. Because most internal documents of ToRP were lost, we cannot know why and when the frequencies were changed now. Hence, in this paper, we describe the history of the NoRP instruments after May 1994.

| | | | 1GHz | 2GHz | 3.75GHz | 9.4GHz |
|---|---|---|---|---|---|---|
| Torii et al. (1979) | LO freq. (MHz) | | 1060 | 2060 | 3732.5 | 9411 |
| | IF freq. (MHz) | | 60 | 60 | 60 | 60 |
| | RF (MHz) | LSB | 1000 | 2000 | 3672.5 | 9351 |
| | | USB | 1120 | 2120 | 3792.5 | 9471 |
| At Nobeyama in May 1994 | LO freq. (MHz) | | 1060 | 2060 | 3810 | 9390 |
| | RF (MHz) | LSB | 1000 | 2000 | 3750 | 9330 |
| | | USB | 1120 | 2120 | - | 9450 |

*Table 1: LO freq. ($f_{Local}$), IF Freq. ($f_{IF}$), and center freq. of RF ($f_{RF}$)*
*in Torii et al, 1979, and at Nobeyama in May 1994.*

1 GHz

From the radio environment survey in the 2000s, we found that the RFI of the 1 GHz observations is caused by signals from aircraft at 1025 -- 1150 MHz. Therefore, we changed the receiver system from a DSB to an SSB to avoid interference. We inserted a broad bandpass filter before the mixer and filtered out the USB on June 1, 2005. Hence, the observing frequency band of the 1 GHz system has been only 1000 MHz ± 5 MHz. This has significantly reduced RFI contamination in the 1 GHz measurements. Because the transponder frequency of a aircraft passed above Nobeyama is not always in the range of 1025 -- 1150 MHz, the interference still occasionally occurs.

On June 25, 2008, the front-end system was replaced with a new system constructed using modern components at the time.

The 1 GHz observation was suspended for a long time because of the mechanical problems of the antenna during the November 29, 2019 – March 27, 2020 and October 10, 2020 – November 14, 2020 periods.

The local oscillator of the 1 GHz receiver system had a glitch between September 2020 and November 8, 2021. Hence the precision of the flux during this period was degraded by approximately 15 % from the nominal value.

2 GHz

On September 6, 1996, the front-end system was replaced with a new system using contemporary components.

In the 2000s, there was a possibility of an increasing in the background level caused by the signal outside the observing frequency range of the 2GHz receiver system. To eliminate this concern, we applied a narrow bandpass filter to the IF signal chain on June 1, 2005.

Because of the overwhelming RFI at 2110 -- 2170 MHz starting on April 12, 2017, the 2GHz observation was suspended. To move the observing frequency to a clean band, we changed the frequency of the local oscillator ($f_{Local}$) to 2120 MHz and placed the bandpass filter before the mixer to filter out the USB. Therefore, the 2GHz receiver was changed to an SSB receiver. The observing frequency of the 2GHz receiver system is 2060 MHz $\pm$ 5 MHz. The observations were resumed on May 15, 2019.

### 3.75 GHz

The original 3.75 GHz receiver system is a DSB receiver, but the system at Toyokawa had already been changed to an SSB receiver with a local frequency of 3810 MHz, before May 1994.

The RFI caused by geosynchronous satellites is an old issue of the 3.75 GHz observation. After a new 3.75 GHz telescope was constructed in 1994, we tuned the observing frequency of the 3.75 GHz receiver system to address this issue. Table 2 presents the frequency-tuning history. In the 1990s, we used a Gunn diode oscillator for a local oscillator at 3.75 GHz. The Gunn diode oscillator has issues with frequency stability and difficulty in changing the frequency. To address these issues, we changed the local oscillator to a synthesizer in September 2006. After the modification, we changed the frequency of the local oscillator ($f_{Local}$) from 3500 MHz to 3820 MHz around the spring and autumn equinoxes, respectively. Owing to change in frequency, the measured solar flux is changed. To reduce this difference, we established a correction factor for each equinox and applied it automatically in the calibration program.

| Date of Change | LO freq. (MHz) | Date of Change | LO freq. (MHz) | Date of Change | LO freq. (MHz) |
|---|---|---|---|---|---|
| 1994/9/26 | 3400 | 2010/3/30 | 3820 | 2015/10/5 | 3500 |
| 1994/9/29 | 3710 | 2010/9/21 | 3500 | 2015/10/26 | 3820 |
| 1994/10/6 | 3750 | 2010/10/25 | 3820 | 2016/2/29 | 3500 |
| 1994 winter | 3825 | 2011/2/21 | 3500 | 2016/3/23 | 3820 |
| 1995/3/8 | 3782 | 2011/2/24 | 3820 | 2016/9/30 | 3500 |
| 1995/5/10 | 3823 | 2011/3/1 | 3500 | 2016/10/26 | 3820 |
| 2003/7/17 | 3762 | 2011/3/31 | 3820 | 2017/2/28 | 3500 |
| 2006/9/21 | 3506 | 2011/10/3 | 3500 | 2017/3/28 | 3820 |
| 2006/10/3 | 3500 | 2011/10/26 | 3820 | 2018/3/1 | 3500 |
| 2006/10/18 | 3820 | 2012/3/1 | 3500 | 2018/3/26 | 3820 |
| 2007/2/21 | 3500 | 2012/3/29 | 3820 | 2018/9/14 | 3500 |
| 2007/3/26 | 3820 | 2012/10/1 | 3500 | 2018/11/15 | 3820 |
| 2007/9/18 | 3500 | 2012/10/22 | 3820 | 2019/2/14 | 3500 |
| 2007/10/22 | 3820 | 2013/3/4 | 3500 | 2019/4/15 | 3820 |
| 2008/2/20 | 3500 | 2013/3/27 | 3820 | 2019/9/17 | 3500 |
| 2008/3/27 | 3820 | 2013/10/3 | 3500 | 2019/11/14 | 3820 |
| 2008/9/24 | 3500 | 2013/10/30 | 3820 | 2020/3/11 | 3500 |
| 2008/10/30 | 3820 | 2014/3/3 | 3500 | 2020/5/20 | 3720 |
| 2009/2/24 | 3500 | 2014/3/26 | 3820 | 2020/7/2 | 3820 |
| 2009/3/23 | 3820 | 2014/10/3 | 3500 | 2020/9/17 | 3500 |
| 2009/10/1 | 3500 | 2014/10/28 | 3820 | 2020/11/13 | 3820 |
| 2009/10/28 | 3820 | 2015/3/3 | 3500 | 2021/1/21 | 3500 |
| 2010/2/22 | 3500 | 2015/3/24 | 3820 | 2021/3/11 | 3820 |

*Table 2: Change log of the local oscillator frequency ($f_{Local}$) in the 3.75 GHz receiver system.*

Around the spring equinox in 2021, we found that the RFI did not diminish even when we changed the frequency of the local oscillator. Hence, we ended the frequency changing to avoid the RFI in the autumn of 2021, and we obtain the 3.75 GHz solar flux from the solar data in the morning because the interference in the morning is weak. Therefore, we, unfortunately, must give up flare observations near spring and autumn equinoxes.

### 9.4 GHz

As mentioned in the previous section, we knew that the marine radars of the ships in the Yaizu port cause interference. To avoid this interference, we tuned the observing frequency of the 9.4 GHz receiver system. In the same way as with the 3.75 GHz system, we used the Gunn diode oscillator for the local oscillator of the 9.4 GHz receiver system. On April 5, 2007, we changed the local oscillator to the synthesizer and fixed the frequency of the local oscillator at 9310 MHz. Owing to the treatment, the occurrence frequency of the interference is significantly reduced.

The polarization selector did not work from September 18, 2014 to June 28, 2021. Hence, the 9.4 GHz telescope was only a radiometer, not a polarimeter, during this period.

Because of amplifier failure, the observation at 9.4 GHz was suspended from June 22, 2017 to December 28, 2017.

The RFI caused by weather radars is currently under investigation. However, interference can be detected in all directions. Hence, it may not be easy to solve this problem.

80 GHz

As mentioned in the early part of this section, the 80 GHz system differs from the other systems of the NoRP and is an interferometer constructed from two 25-cm parabolic antennas. Even at the Nobeyama highland, the effect of the atmosphere cannot be neglected at such a higher observing frequency. Hence, the 80 GHz system was constructed as the interferometer for removing the atmospheric effects (nulling interferometer). It means that the baseline length of the antennas is fixed for removing the component of quiet sun. Therefore, we cannot obtain the total solar fluxes with the 80 GHz system, and the data obtained with the system reveals the 80 GHz fluxes from only bright and small solar phenomena, like big solar flares (larger than GOES X-class, or very radio-loud flares). The details of the 80 GHz system in Nakajima et al. (1985).

Owing to the failure of the polarization selector, the 80 GHz telescope lost the function of measuring the polarization degree and became just a radiometer on June 23, 2005. Unfortunately, we do not have any plan for repairing it at present.

The gearbox of the antenna that mounts the parabolic dishes for 35 and 80 GHz observations broke on November 25, 2018, and the observations with these frequencies were suspended. They were resumed on October 15, 2021.

We have not performed any significant modifications to the 17 and 35 GHz receiver systems since May 1994.

It is not essential to understand the NoRP datasets, but we note that the back-end systems for 1, 2, 3.75, 9.4, and 17 GHz were replaced as follows: on December 4, 2002 for 1 and 2 GHz, on October 7, 2003 for 3.75 GHz, on December 19, 2003 for 9.4 GHz, and on March 8, 2004 for 17 GHz.

## 4. Observation data obtained with Toyokawa and Nobeyama Radio Polarimeters and their metadata

All digitized observation data obtained with the ToRP and NoRP were archived on the Solar Data Archive System (SDAS)[5] operated by the Astronomy Data Center of the NAOJ, and the data can be obtained on the Internet from the website and anonymous ftp site of the NoRP. In this section, we introduce the available datasets and their properties.

### Raw data files

In the back-end system of the NoRP and ToRP, the power of the difference between the solar signal and the reference signal generated in the front-end system is measured. The solar signals with the right (R) and left (L) circular polarizations are included in the IF signal with time sharing, and each signal is integrated for 100 ms. Then, the back-end system outputs the integrated summing (R+L), and difference (R-L) signals as the voltages. The raw data of the NoRP and ToRP stored in the digital files consist of the 12-bit signed numbers that are converted from these voltages using an A/D converter.

The digitization of the ToRP was done in early 1979 (Shibasaki, Ishiguro, and Enome 1979), but the digital data obtained with the ToRP only since February 1988 are stored in the SDAS. On the other hand, the 17, 35, and 80 GHz receiver systems in Nobeyama were digitized in March 1990. Their data files are stored in the SDAS and can be obtained from the following URLs: "https://solar.nro.nao.ac.jp/norp/raw/" and "ftp://solar-pub.nao.ac.jp/pub/nsro/norp/raw/." One data file includes the data obtained in a day. The filenames of the ToRP and NoRP are "tyYYMMDD.XXX" and "plYYMMDD.dat," respectively. "YYMMDD" is the observation date in Japan Standard Time (offset: UT +9 hours), and "XXX" indicates the observing frequency, Stokes, and name of the status data. After constructing the NoRP from the ToRP and the 17, 35, and 80 GHz antennas at Nobeyama, the data with all the observing frequencies are stored in a file "plYYMMDD.dat."

[5] https://hinode.nao.ac.jp/SDAS/index_e.shtml

The data archive is not searchable, but is organized by year and month such that the data for a given date are readily found.

The ToRP/NoRP raw data files have distinctive formats and can be read using only the Interactive Data Language (IDL) with the Solar SoftWare package (SSW: Freeland and Handy, 1998 ). The ToRP/NoRP data reader implemented on the SSW calibrates the data using the methods described in Tanaka et al. (1973) and Nakajima et al. (1985). The data analysis guide for the NoRP using the IDL and SSW is provided at the following URL: "https://solar.nro.nao.ac.jp/norp/doc/man_nroe.pdf".

### Calibrated data files with the XDR format

The data calibrated with the ToRP/NoRP reader on the SSW are also stored on the SDAS. The URL of the datasets is "https://solar.nro.nao.ac.jp/norp/xdr/" and "ftp://solar-pub.nao.ac.jp/pub/nsro/norp/xdr/". The calibrated data is filed using the "eXternal Data Representation" format (XDR: RFC4506[6]). The data file is generated automatically after the day's observation and is then released to the public immediately after the file generation. The directory structure is the same as that of the raw data.

XDR is one of the standard data serialization formats, but we do not check which files can be read using programs other than IDL. Hence, use of the XDR files generally requires access to IDL.

### Calibrated data files with the FITS format

The IDL + SSW environment is the de fact standard for solar data analysis. When a new instrument for solar observations, for example, the solar telescope aboard a satellite is developed, most developing teams provide the data analysis software for the instrument in the IDL + SSW environment. Hence, the two data files (raw and xdr files) described in the previous sections are sufficient for the investigatation of microwave data by solar physicists because solar researchers already have an environment. However, solar microwave data are helpful for studies in other fields, such as those on the Earth's upper atmosphere, climate, and exo-planets. To encourage the use of NoRP data for such studies, it is not helpful that the IDL + SSW environment is required for NoRP data analysis because IDL is commercial software.

---

[6] https://www.ietf.org/rfc/rfc4506.html

We have therefore created the NoRP data file from the raw data obtained in June 1994 using the Flexible Image Transport System (FITS) format (IAU FITS Working Group, 2016) to solve this problem. The FITS format is the most widely used format for the data transfer of astronomical data, and there are libraries for manipulating the files in this format in many environments, for example, FORTRAN, C, Python, Java, and Perl. Hence, anyone can use NoRP data for research without commercial software when NoRP FITS files are available. To encourage the NoRP data analysis in other fields, we created the NoRP FITS file database in 2021.

A NoRP FITS file includes raw data, data calibrated with the NoRP reader of IDL automatically, and instrument status for each day. The URLs for accessing the database are "https://solar.nro.nao.ac.jp/norp/fits/" and "ftp://solar-pub.nao.ac.jp/pub/nsro/norp/fits". The filename of a NoRP FITS file is "norpYYMMDD.fits.gz", and the directory structure is the same as that for the raw data files. The details of NoRP FITS files are described in the following URL "https://solar.nro.nao.ac.jp/norp/html/fits_png_ql/ReadMe.pdf".

To improve the visibility of the data, we have created quick-look webpages of the NoRP data based on the NoRP FITS files (https://solar.nro.nao.ac.jp/norp/html/fits_png_ql/). From the webpage, one can access the one-day time profile of solar Stokes I and Stokes V signals with all the observing frequencies of the NoRP, as shown in Figure 9. In addition, we have built a Python library for reading NoRP FITS files; this library can be obtained from the URL: "https://solar.nro.nao.ac.jp/norp/Python/norp_fits.py".

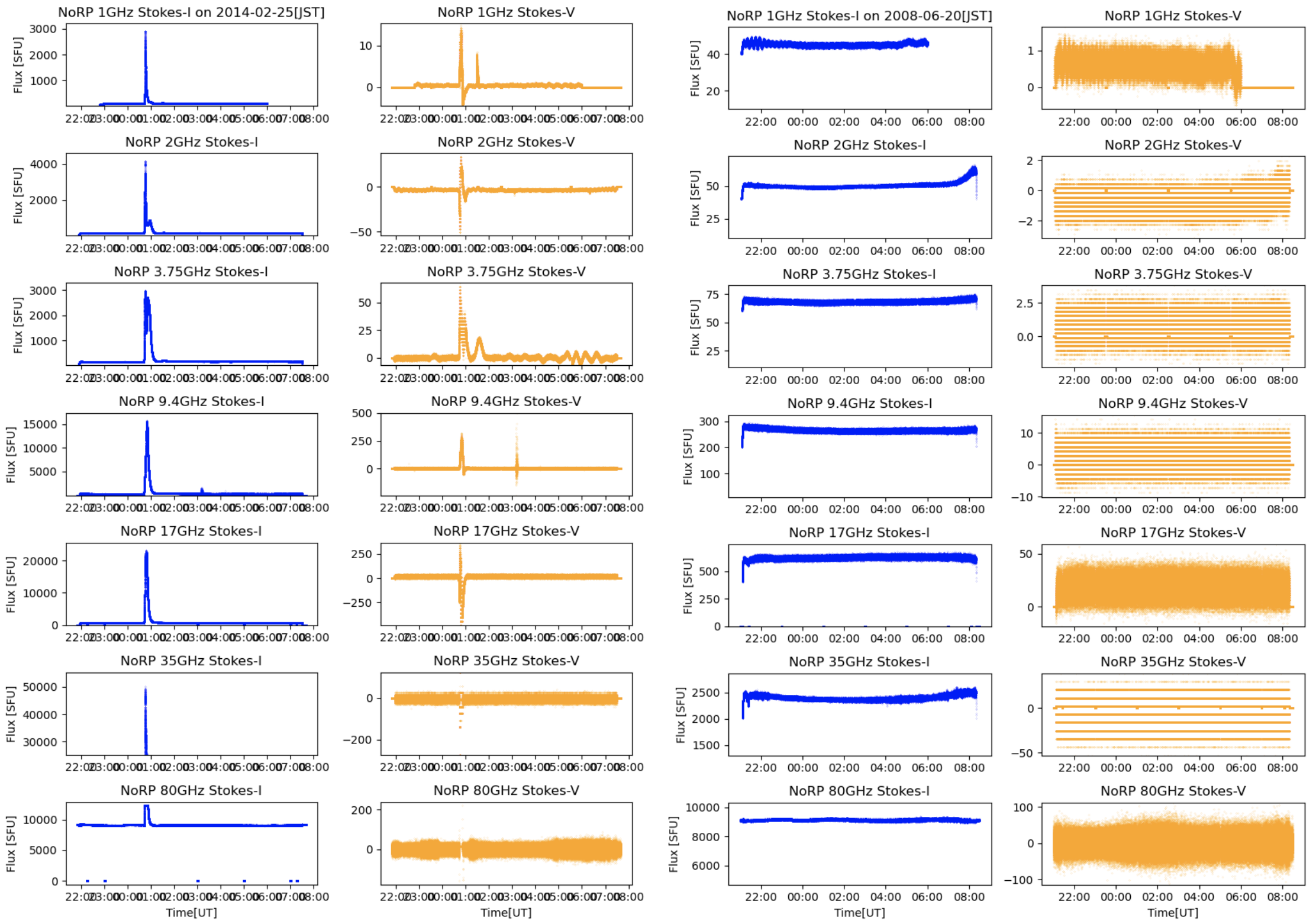

*Figure 9: Example of the quick-look plots from the NoRP FITS files.*

*From the left, the time profiles of Stokes-I (Blue) and Stokes-V (Yellow) on February 25, 2014 and June 20, 2008. A radio burst associated with an X4.9 class flare was observed on Feburary 25, 2014.*

## Table and CSV file of daily solar fluxes

Measurement of the daily solar total flux using the ToRP and the 17 GHz polarimeter at Mitaka began immediately after their monitoring observations started, and the daily solar total fluxes were published in QBSA as mentioned in Section 2. Calibration was performed manually using a chart paper and ruler before digitalizing the instruments. After moving the ToRP to Nobeyama and constructing the NoRP, we continued to measure the daily solar fluxes at 1, 2, 3.75, 9.4, and 17 GHz. We continued manual calibration even after the digitization of the instruments was completed. Obviously, we used computers instead of chart paper and rulers, but we manually selected the time range of the values for calculating solar fluxes. Time ranges with no radio bursts and no RFI are selected for the calculation. The importance of manual selections has grown now because the influence of RFI has been worsening since the 2000s.

The publication of the QBSA was terminated in 2009 because of the web-based form of information exchange. We digitized the data tables of the ToRP, 17 GHz polarimeters at Mitaka, and NoRP used for reporting at QBSA and stored them in the SDAS. One can get the text files of the one-month daily fluxes tables since 1951 from "https://solar.nro.nao.ac.jp/norp/data/daily/". The filename of the ToRP, Mitaka, and NoRP is "tykwYYMM", "tokYYMM", and "nbymYYMM", respectively. "YYMM" indicates the year and month of the table.

Although the total solar fluxes at 17 GHz are recorded in the text files, the values are more affected by the atmosphere variation above the Nobeyama area than the data at the lower frequencies. Hence, we should treat the data carefully when the 17 GHz data is used to research long-term solar variations.

Text files are inconvenient for manipulating the data. Hence, we prepared the CSV file that includes the total solar fluxes at 1, 2, 3.75, and 9.4 GHz observed with ToRP and NoRP since November 1951. The CSV file can be obtained from " https://solar.nro.nao.ac.jp/norp/data/daily/TYKW-NoRP_dailyflux.txt".

Note the values in the text files and the CSV file. The values recorded in these files are the total flux densities measured on the Earth. The distance between the Sun and Earth depends on the season so that plots show he seasonal variation, as in the upper panel of Figure 10. This choice means that the archive data are directly suitable for geophysics studies. On the other hand, for solar studies, the users should revise the data with the distance between the Sun and Earth, as shown in the lower panel of Figure 10.

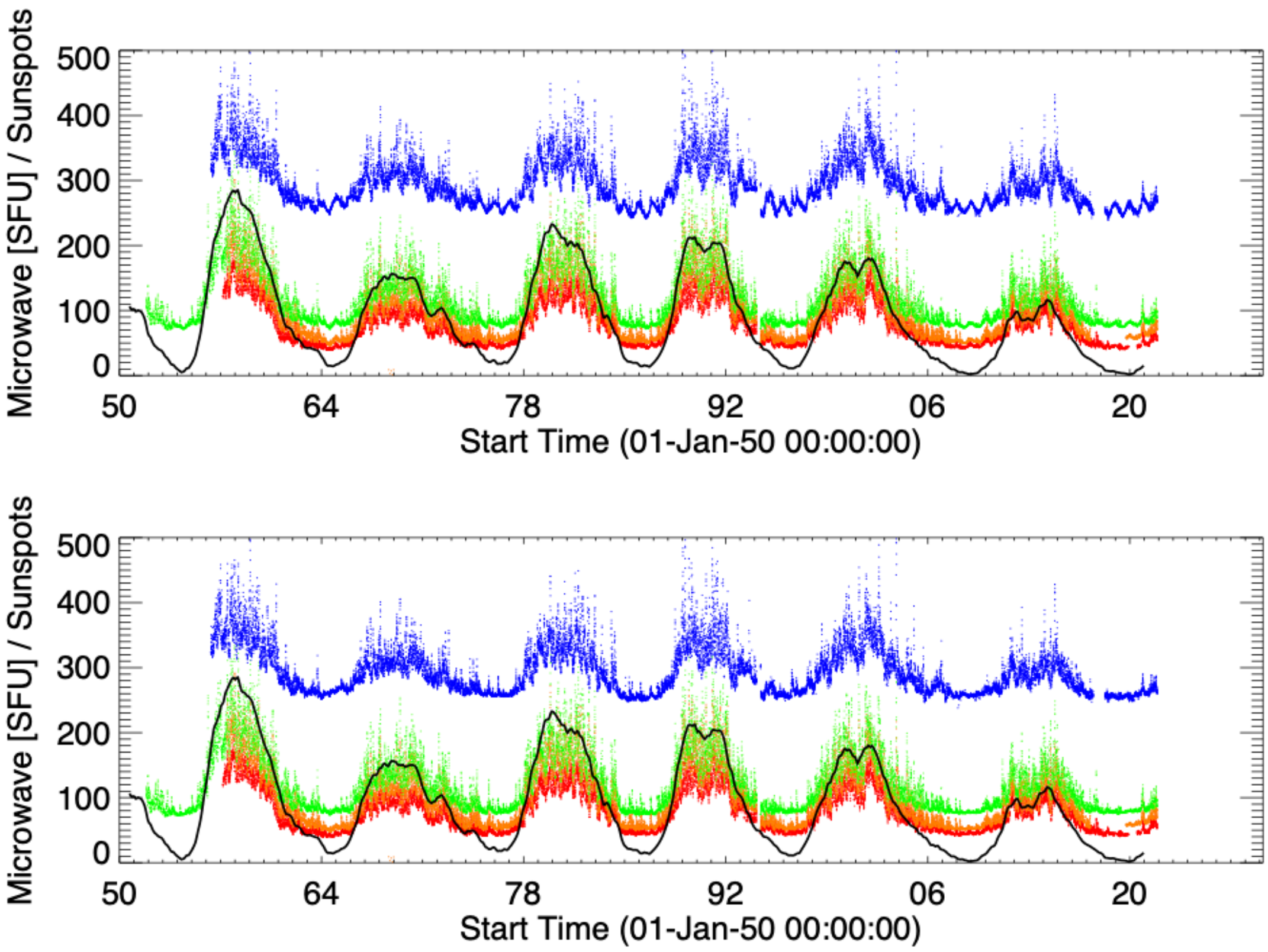

*Figure 10: Long-term variations of microwave fluxes and sunspots from the 1950s to December 2021. The upper panel shows the original data, and the lower panel presents the revised data using the distance between the Sun and Earth. Red: 1GHz, orange: 2 GHz, green: 3.75 GHz, blue: 9.4 GHz, and black: 13-months smoothed sunspot number compiled by SILSO, Royal Observatory of Belgium, Brussels.*

Shimojo et al. (2017) used the datasets and proposed a new method to determine the epoch of solar maxima and minima based on the monthly standard deviation of the microwave fluxes. They also displayed the microwave spectra at the solar maxima and minima from Cycle 19 to Cycle 24, defined using their method. We incorporate their microwave spectra, adding the Cycle 25 solar minimum spectrum, as examples of the solar microwave spectra obtained with the NoRP.

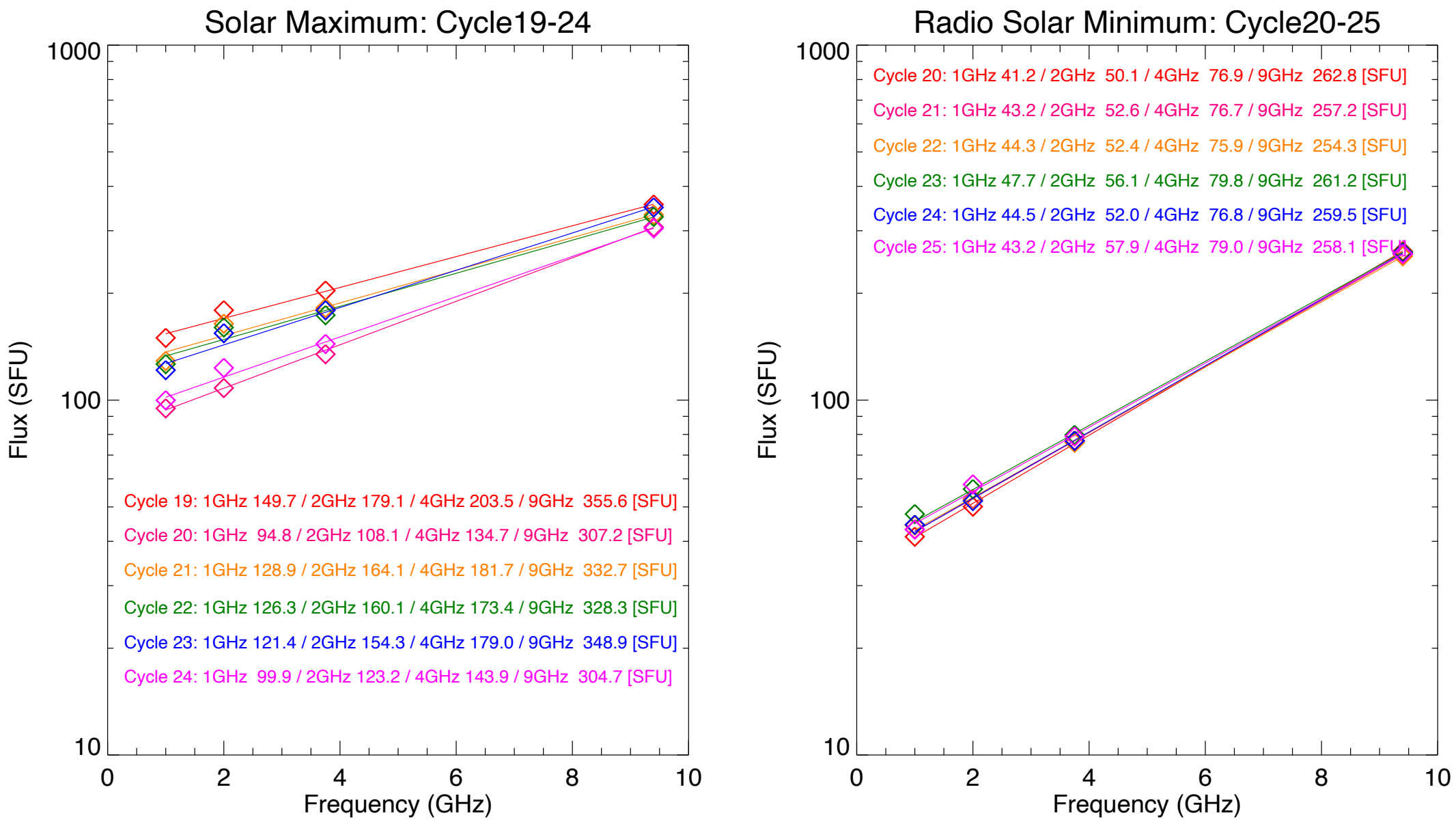

*Figure 11: Monthly mean microwave spectra in the months of the solar maxima (left panel) and the solar minima (right panel). The months of the spectra are determined using the method described in Shimojo et al. (2017). The colors indicate the number of the solar cycle. The solid lines are the fitting result using a linear function for each spectrum.*

### List, time profile, and calibrated data of radio bursts

The NoRP web page contains a list of solar radio bursts from 1988. The URL of this list is "https://solar.nro.nao.ac.jp/norp/html/event/". The list includes the link to an individual event (Figure 11), and the time profile of the bursts is presented on the event page. The event page also includes a link to the calibrated data file with XDR format.

The events since 1992 had been detected using the correlation data obtained with the Nobeyama Radioheliograph (NoRH: Nakajima et al. 1994). Hence, the list after 1992 includes only the events that occurred during the observing period of the NoRH. The NoRH observing period of one day is shorter than NoRP. So, the events that occurred in the early morning (before 23:00UT) or late afternoon (after 6:30UT) in Japan do not include in the list even when NoRP operated. Before starting the NoRH observations, the events had been detected by visual inspection from the charts.

As mentioned, the event detection utilizes a tight coupling between the NoRH and NoRP via a computer network. However, after the operation of the NoRH was moved from the NAOJ to Nagoya University on April 1, 2015, we could not establish a tight coupling due to network security issues. Hence, the list includes only the events that occurred before March 31, 2015.

# Nobeyama Radio Polarimeters Event 20050120_0638

to Event List, to XDR (IDL save) Data, to Nobeyama Radio Observatory

---

EventID : 20050120_0638
Start : 2005-01-20T06:38:40.000
Peak : 2005-01-20T06:46:00.000
End : 2005-01-20T07:00:00.000
Maximum flux I @ 1GHz (SFU) : 0
Min/Max flux V @ 1GHz (SFU) : 0/0
Maximum flux I @ 2GHz (SFU) : 11819
Min/Max flux V @ 2GHz (SFU) : -2706/77
Maximum flux I @ 3.75GHz (SFU) : 29553
Min/Max flux V @ 3.75GHz (SFU) : -2069/864
Maximum flux I @ 9.4GHz (SFU) : 61360
Min/Max flux V @ 9.4GHz (SFU) : -1496/263
Maximum flux I @ 17GHz (SFU) : 87174
Min/Max flux V @ 17GHz (SFU) : -3806/-18
Maximum flux I @ 35GHz (SFU) : 90992
Min/Max flux V @ 35GHz (SFU) : -4870/0
Maximum flux I @ 80GHz (SFU) : 2077

---

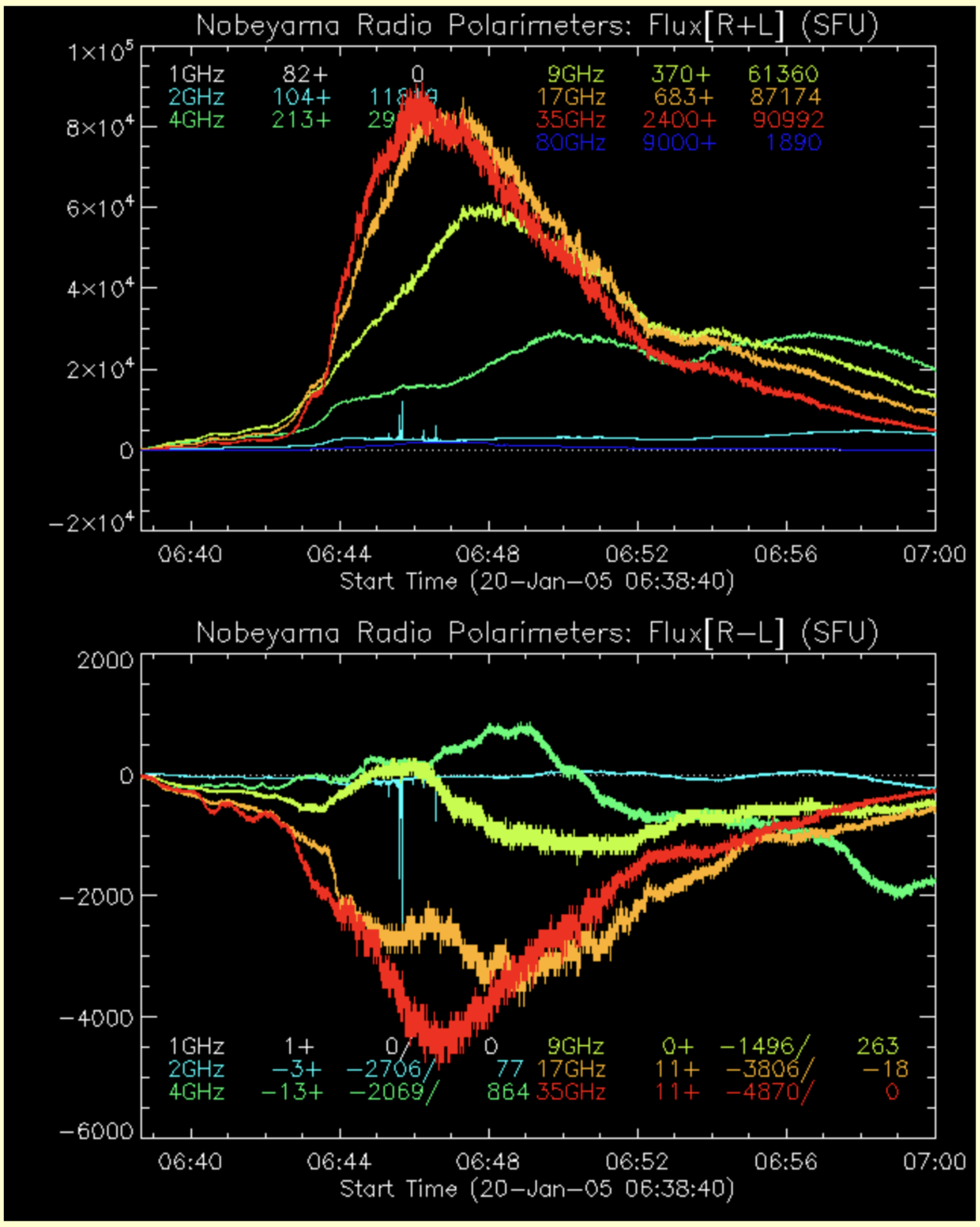

---

to Event List, to XDR (IDL save) Data, to Nobeyama Radio Observatory

*Figure 12: An example of the event page: a radio burst associated with an X7.1 flare on January 20, 2005*

## 5. Summary and the future of the Nobeyama Radio Polarimeters

In the paper, we have briefly described the history of the solar monitoring observations in the microwave range in Japan, the transition of the NoRP receiver systems after May 1994. This paper has also described the datasets archived and provided by the NAOJ and their properties. At present (April 2022), all antennas and receiver systems of the NoRP are working well, and we measure solar fluxes at 1, 2, 3.75, 9.4, 17, and 35 GHz and their polarization every day and large flare fluxes with 80 GHz. Nevertheless, the current operation of NoRP is a continuous struggle against deterioration; the oldest antenna has been in service for over 40 years. We replace parts of antennas and receivers with modern instruments, but long suspensions of service have frequently occurred after the 2010s. They are caused by the lack of human resources for operation and lost information of the ToRP and NoRP. In long-term observations, it is challenging to maintain the history of instruments, when most information is recorded on paper media. We hope that this paper will be helpful for future users of the NoRP data.

In section 4, we reviewed the digitized ToRP and NoRP data. Before the digitization, the raw data of these instruments were recorded in chart papers. Chart papers that recorded the ToRP data from 1958 to 1979 were microfilmed in the 1980s, and had been archived in Nobeyama. We are working on digitizing the microfilmes.

As mentioned in Section 2.3, the NoRP operation has been under the 5-years extended operation period since April 2019. We will make efforts to continue the observation with NoRP.

## Acknowledgement

The authors would like to thank all the people who have participated in the operation of the Toyokawa and Nobeyama Radio Polarimeters for over 70 years. We thank Prof. Hiroshi Nakajima and Prof. Kiyoto Shibasaki. They readily consented to review the paper and sent us useful comments. We also thank Prof. Masato Ishiguro for providing a photograph of ToRP. Moreover, we thank Dr. Hisashi Hayakawa and the reviewers for their useful comments. The Nobeyama Radio Polarimeters (NoRP) are operated by the Solar Science Observatory, a branch of the National Astronomical Observatory of Japan, and their observing data are verified scientifically by the consortium for NoRP scientific operations. This work was carried out on the Multi-wavelength Data Analysis System (MDAS) and Solar Data Archive System (SDAS) operated by the Astronomy Data Center of the National Astronomical Observatory of Japan. The authors thank the SILSO, Royal Observatory of Belgium for providing International Sunspot Number.